\documentclass[aps,prd,nofootinbib,floatfix,amssymb]{revtex4} 
\usepackage{epsf,feynmf}
\usepackage{graphicx,epsfig}
\usepackage{amsfonts}
\usepackage{amssymb}

\def\gsim{\mathrel {\vcenter {\baselineskip 0pt \kern 0pt \hbox{$>$} \kern 0pt \hbox{$\sim$} }}}

\def\lsim{\mathrel {\vcenter {\baselineskip 0pt \kern 0pt \hbox{$<$} \kern 0pt \hbox{$\sim$} }}}
\newcommand{\Vt}{{\tilde V}}

\newcommand{\dpc}{\delta \phi_c }
\newcommand{\dpq}{\delta \phi_q}
\newcommand{\bdv}[1]{{\bf{#1}}}
\newcommand{\Q}[2]{Q_{\bdv{#1}}^{#2}}

\newcommand{\U}[2]{U_{#1}^{#2}}
\newcommand{\Ud}[2]{U_{#1}^{#2}{}^{*}}
\newcommand{\dU}[2]{\dot{U}_{#1}^{#2}}
\newcommand{\dUd}[2]{\dot{U}_{#1}^{#2}{}^{*}}
\newcommand{\Qp}[2]{Q_{\bdv{#1}}^{#2}{}^{+}}
\newcommand{\Qm}[2]{Q_{\bdv{#1}}^{#2}{}^{-}}

\newcommand{\qdp}{\bdv{q}\cdot\bdv{k}}

\begin{document}

\title{Cosmology With Many Light Scalar Fields: Stochastic Inflation and Loop Corrections}
\author{Peter Adshead}
\author{Richard Easther}
\affiliation{Department of Physics\\ Yale University, New Haven, CT 06511 USA}
\author{Eugene A. Lim}
\affiliation{Department of Physics and ISCAP,  \\ Columbia University, New York, NY 10027 USA}

 \begin{abstract}
We explore the consequences of the existence of a very large number of light scalar degrees of freedom in the early universe.    We distinguish between {\em participator\/} and {\em spectator\/} fields. The former have a small mass, and can contribute to the inflationary dynamics; the latter are either strictly massless or have a negligible VEV.    In N-flation and generic assisted inflation scenarios, inflation is a co-operative phenomenon driven by $N$ participator fields, none of which could drive inflation on its own.  We   review upper bounds on $N$, as a function of the inflationary Hubble scale $H$.  We then consider stochastic and eternal inflation in models with $N$ participator fields showing that individual fields may evolve stochastically while the whole ensemble behaves deterministically, and that a wide range of eternal inflationary scenarios are possible in this regime. We then compute one-loop quantum corrections to the inflationary power spectrum. These are largest with $N$ spectator fields and a single participator field, and the resulting bound on $N$ is always {\em weaker\/} than those obtained in other ways. We find that loop corrections to the N-flation power spectrum do not scale with $N$, and thus place no upper bound on the number of participator fields. This result also implies that, at least to leading order, the theory behaves like a composite single scalar field.  In order to perform this calculation, we address a number of issues associated with loop calculations in the Schwinger-Keldysh ``in-in'' formalism.

\end{abstract}

\maketitle
 
%-------------------------------------------------------------------------------------------------------------
%-------------------------------------------------------------------------------------------------------------
\section{Introduction}
%-------------------------------------------------------------------------------------------------------------
%------------------------------------------------------------------------------------------------------------- 
 
Many candidate theories of fundamental physics predict the existence of large numbers of scalar degrees of freedom.  Classically, if these modes are not excited they play no role in the cosmological dynamics.  Quantum mechanically, however, light scalar modes fluctuate with an amplitude set by the Hubble scale $H$ and, since everything couples to the graviton, they contribute to loop corrections. Thus, while adding $N$ light scalar modes need not change the classical dynamics of the early universe, we expect quantum  contributions that scale with $N$.  In addition, if these fields are in thermal equilibrium with the rest of the universe their contribution to the effective number of degrees of freedom modifies the relationship between density and temperature.  For a given $H$ we can therefore find an upper limit on  $N$, above which these modes dominate the cosmological evolution.

Consider a scenario with $N$ scalar fields,
\begin{equation}
S =  \int d^4 x \sqrt{g}\left[\frac{M_p^2}{2}R+ \sum_I \left(-\frac{1}{2}(\partial{\phi_I})^2 + V_I(\phi_I)\right) \right] \, ,\label{eqn:Nflationaction}
\end{equation}
where the total potential $V = \sum_I V_I(\phi_I)$ is a sum of $N$ uncoupled potential terms.\footnote{We use the reduced Planck mass $M_p^2 = 1/8\pi G$ throughout.}  A field  $\phi_I$  is ``light''  if   $d^2V_I/d\phi_I^2  /2\equiv m_I^2 \ll H^2$.  We make a distinction between {\em participator\/} and {\em spectator\/} fields.  The latter are either massless because $V_I$ is strictly zero, or their vacuum expectation value (VEV) is very small. Conversely, participator fields have small but non-zero mass and  sufficient VEV for them to help drive inflation via their contribution to the overall density. 

Perhaps the simplest route to a meaningful bound on $N$ is to  note that all these fields undergo quantum mechanical fluctuations of order $\delta \phi_i \sim H/2\pi$. During inflation these fluctuations freeze out as classical perturbations at scales larger than the Hubble length,  $1/H$.  Each field has gradient energy  $(\nabla \phi)^2/2$, which counts towards the overall energy density.  The  gradient energy thus scales like $N(\delta \phi /\delta x)^2 /2 \sim N H^4/8 \pi^2$.  Given $H$, the energy density is provided by the $0-0$ Einstein equation, $H^2 = \rho/3{M_p^2}$.  For self-consistency,  the gradient  contribution must be much smaller than other contributions to $\rho$, so
\begin{equation}
N \ll \frac{M_p^2}{H^2}. \label{eqn:gradientbound}
\end{equation}
If  inflation occurs at the GUT scale, then $M_p/H\sim 10^5$ and $N\ll 10^{10}$. This bound can be derived by many routes  (e.g. \cite{Huang:2007zt,Dvali:2008sy}), and appears to be robust.\footnote{In \cite{Ahmad:2008vy,Ahmad:2008eu} this limit on $N$ is derived in the context of N-flation but in reality it applies to {\em any\/} scenario with $N$ light fields.}

One can also consider bounds on $N$ from loop corrections to the gravitational constant. Since gravity (and thus the graviton) couples to all fields, matter loop corrections can renormalize its value \cite{Zee:1981mk,Adler:1980bx}. Veneziano \cite{Veneziano:2001ah} argued that in order for the effective value of Newton's constant to be greater than zero we need
\begin{equation}
N \ll \frac{M_p^2}{\Lambda^2}, \label{eqn:speciesbound}
\end{equation}
where $\Lambda$ is the scale of the invariant UV cut-off, for example the mass scale of the $N$ stable fields.  Likewise, Veneziano \cite{Veneziano:2001ah} points  out that this bound prevents potential violations of the holographic bound on the entropy density that can be encountered when the total number of species grows without limit   \cite{Unruh:1982ic, Sorkin:1981wd}, the so-called ``Species Problem''. More recently, Dvali \cite{Dvali:2007hz} noted that in the presence of a large number of species, the Veneziano mechanism weakens the gravitational coupling by a factor of $1/N$.  Setting  $N\sim 10^{32}$  and working at the TeV scale to satisfy the bound of equation~(\ref{eqn:speciesbound})  solves the  hierarchy problem, provided some ultraviolet completion of the standard mode can produce the requisite  value of $N$.  Dvali also provides an alternate non-perturbative derivation of equation (\ref{eqn:speciesbound}) based on the consistency of black hole physics. 

In Section~ \ref{sect:eternal} we begin by  considering stochastic inflation \cite{Vilenkin:1983xq,Linde:1986fd,Guth:2000ka} with multiple degrees of freedom. Many simple inflationary potentials possess a range of field values for which the potential is safely sub-Planckian, while the stochastic motion of the field  dominates the semi-classical rolling.  With a single field, a stochastic phase is necessarily eternal, since the inflationary domains perpetually reproduce themselves. However, once $N>1$, more complicated scenarios become possible.  Specifically, with  $N$ participator  fields we can implement {\em assisted\/} eternal inflation without any single field acquiring a super-Planckian VEV. This scenario is a variety of ``assisted inflation'' where the inflaton is a composite of many individual fields  \cite{Liddle:1998jc}.  Secondly, we find solutions where {\em individual\/} fields move stochastically, but  a well-defined composite field rolls smoothly towards its minimum. Finally, we find models where a field (or fields) evolves semi-classically, along with other fields that move stochastically. If these fields have symmetry breaking potentials they yield an apparent ``multiverse'' of disconnected bubbles. However, the stochastic phase ends globally, as the semi-classically evolving field gives a natural cut-off to what would otherwise be an eternally inflating universe, providing a potential toy model for studying the well-known measure problem in eternal inflation \cite{Linde:1993xx,Guth:2000ka,Garriga:2005av,Easther:2005wi, Bousso:2006ev, Aguirre:2006ak,Easther:2007sz}.  

All fields couple to the inflaton gravitationally, and thus contribute loop corrections to the inflation potential, via a coupling of order $(H/M_p)^2$. This is necessarily a small number during the last 60 e-folds of inflation, given that $H$ fixes the scale of tensor fluctuations and is bounded  by the absence of an observed B-mode in the microwave background.  Consequently, any single loop makes a tiny  correction to the  inflaton propagator. However, this contribution is amplified by $N$, the number of species that can flow round the loops, and in Section \ref{sect:loop} we compute the relevant one-loop corrections to the inflaton propagator.   We consider two limits -- $N$ spectator fields with a single inflaton, and the N-flation case, with $N$ participator fields  \cite{Dimopoulos:2005ac,Easther:2005zr}.  With $N$ spectator fields we find an upper bound on $N$  similar to those of  \cite{Veneziano:2001ah, Dvali:2007wp}.    We describe these results in  Section \ref{sect:loop}, relegating many details to Appendix \ref{app:1loop}. 
We work in the ``in-in" formalism \cite{Schwinger:1961,Keldysh:1964ud,Jordan:1986ug,Calzetta:1986ey}, which has been turned into an extremely powerful tool for studying higher order corrections to cosmological correlations by Maldacena \cite{Maldacena:2002vr}  and  Weinberg \cite{Weinberg:2005vy}.  This calculation requires us to develop the fourth order interaction Hamiltonian for a theory of inflation with $N$ uncoupled scalar fields,  which we present in Appendix \ref{App:4ptderivation}, and which will have applications beyond the current calculation. Moreover, it turns out that there are some subtleties with the computation  of loop corrections in the in-in formalism which we also clarify in the Appendices.   Finally we conclude in Section \ref{sect:conclusions}.

%-------------------------------------------------------------------------------------------------------------
%-------------------------------------------------------------------------------------------------------------
\section{Stochastic $N$-flation} \label{sect:eternal}
%-------------------------------------------------------------------------------------------------------------
%-------------------------------------------------------------------------------------------------------------

In simple versions of $N$-flation,  one has $N$ identical participator fields, and inflation emerges as a cooperative phenomenon.   We assume that the overall potential is the sum of the  individual fields' potential terms, and that cross terms are absent.  Each field only feels the ``slope'' of its own potential, but the corresponding friction term in the field's equation of motion is still proportional to $H$, and thus grows with $N$.  Interestingly, a similar scaling also applies to {\em stochastic\/} inflation, which occurs when the inflaton evolution is dominated by  quantum fluctuations, rather than semi-classical rolling \cite{Vilenkin:1983xq,Linde:1986fd,Guth:2000ka}.  The individual fields have fluctuations of order $H$, so this amplitude will grow relative to the semi-classical evolution as $N$ is increased.

Let us begin by looking at $N$ fields with identical potential terms and initial VEVs.   Assuming slow roll,
\begin{equation}
H^2 = \frac{1}{3M_p^2} \sum_I V_I(\phi_I) =  \frac{1}{3M_p^2}  N \Vt,  \label{eq:Hgeneric}
\end{equation}
where $\Vt = V_I(\phi_I)$ is the potential for any one of the fields.
The amplitude of the quantum fluctuations of the $I-$th field is 
\begin{equation}
\delta \phi_{I,q}  = \delta \phi_q = \frac{H}{2 \pi},
\end{equation}
where the first equality reflects our assumption of identical fields. From  equation (\ref{eq:Hgeneric})  
\begin{equation}
\delta \phi_q   = \sqrt{\frac{1 }{12\pi^2}  \frac{N\Vt}{M_p^2}} \, .
\end{equation}
Conversely, the distance travelled by $\phi_I$ in a single Hubble time is
\begin{equation}
\delta \phi_c   =  |\dot{\phi_I}| \frac{1}{H} =  \frac{V'_I}{3H}   \frac{1}{H}  =  \frac{M_p^2 V'_I}{ N \Vt},
\end{equation}
where $V'_I = \partial V/\partial \phi_I = \partial V_I/\partial \phi_I$.
The defining condition for the stochastic inflation is that  $\delta \phi_c < \delta \phi_q$ \cite{Vilenkin:1983xq,Linde:1986fd}. Forming the ratio of these terms gives
\begin{equation}
\frac{\dpq}{\dpc} =   \sqrt{\frac{1}{ 12 \pi^2}}\left(\frac{ N\Vt}{M_p^2}\right)^{3/2} \frac{1}{V'_I},
\end{equation}
and stochastic inflation therefore occurs whenever the above expression is larger than unity.  This ratio increases with $N$ if everything else is held fixed.  This is to be expected, since boosting $N$ lifts the overall energy density, and thus the amplitude of the quantum fluctuations. Conversely, the semi-classical rolling slows with $N$, as we are increasing $H$ while holding $V_I'$ constant.  

The above discussion holds for a generic potential, but now consider  $N$  quadratic potentials with identical masses, $V_I = m^2 \phi_I^2 /2$. Assisted inflation can be described in  terms of an effective single field, $\varphi$ which, for quadratic potentials  is $\phi_I = \varphi/\sqrt{N}$ and \cite{Dimopoulos:2005ac,Easther:2005zr}
\begin{equation}
\frac{\dpq}{\dpc} =    \sqrt{ \frac{ N}{96\pi^2}}  \frac{N \phi_I^2}{M_p^2}\frac{m}{M_p}=
   \sqrt{ \frac{ N}{96\pi^2}}  \frac{\varphi^2}{M_p^2}\frac{m}{M_p}.\label{eqn:stochasticcond}
\end{equation}
 Interestingly, this expression {\em retains\/} an explicit dependence on $N$, whereas N-flation ends at a fixed value of $\varphi$, independently of $N$. Before proceeding, let us consider some specific numbers.  The principal virtue  of N-flation is that we can build a GUT-scale inflation scenario without invoking trans-Planckian VEVs. Consequently, if  $\phi_I \sim M_p$  the critical value of $N$ at which the fields can move stochastically is
\begin{equation}
N = (96 \pi^2)^{1/3} \left(\frac{M_p}{m} \right)^{2/3} \approx 10 \left(\frac{M_p}{m} \right)^{2/3} \, .
\end{equation}
The ratio $m/M_p$ is fixed via the amplitude of the perturbation spectrum, and is around $2 \times 10^{-4}$ \cite{Dimopoulos:2005ac,Easther:2005zr}.   Consequently, we might see the onset of stochastic motion in the fields if $N \sim {\cal{O}}(10^4)$.    This is an order of magnitude larger  than the number of fields one expects in N-flation, on the basis of the likely number of two-cycles (from which the N axions are derived) one can find in a realistic Calabi-Yau \cite{Dimopoulos:2005ac,Easther:2005zr}, but is much less than the absolute upper limit on $N$ given by equation (\ref{eqn:gradientbound}).   To satisfy this requirement, we need  $H^2 \approx N m^2/ 6$ (assuming $\phi_I = M_p$), in which case 
\begin{equation} \label{eqn:Nmax}
N\lsim \sqrt{6}\frac{M_p}{m} \sim 10^5 . 
\end{equation}
The theoretical upper limit on $N$ is roughly an order of magnitude larger than the value required for the individual fields to move stochastically.\footnote{Huang et. al. \cite{Huang:2007zt,Huang:2007st} has argued that choatic eternal inflation with large $N$ fields is ruled out by the so-called ``weak gravity conjecture'' introduced in \cite{ArkaniHamed:2006dz}.} Conversely, slow roll ends when  $\varphi \approx \sqrt{2}M_p $ or $\phi \approx \sqrt{2/N}M_p$. For self-consistency, we expect the $N$ fields to be moving semi-classically at this point, so we can derive a very weak upper bound on $N$ by setting $\varphi \approx \sqrt{2}M_p$ in  (\ref{eqn:stochasticcond}), namely $N \lsim 24\pi^2 M_p^2/m^2 \sim 10^{10}$, far above even the weak bound of equation (\ref{eqn:gradientbound}).  Since the last sixty e-folds occur at  values of $\varphi$ a few times larger than $\sqrt{2}M_p$, we need not worry that the fields are moving stochastically over astrophysically interesting scales.  

With a single field, the onset of stochastic inflation is synonymous with the amplitude of the density fluctuations exceeding unity \cite{Creminelli:2008es}. For $N$-flation the density fluctuations have an amplitude \cite{Dimopoulos:2005ac,Easther:2005zr}
\begin{equation}
P_{\cal R}^{1/2} = \sqrt{ \frac{ 1}{96\pi^2}}  \frac{\varphi^2}{M_p^2}\frac{m}{M_p},
\end{equation}
which is $\sqrt{N}$ {\em smaller\/} than the critical ratio for the onset of stochastic motion by the individual fields, so at the threshold for stochastic motion, $P_{\cal R}^{1/2} \sim 1/\sqrt{N} \ll 1$.   In this case {\em stochastic\/} inflation is not synonymous with {\em eternal\/} inflation.  As described above, in one Hubble time each field makes a random jump of $\sim \delta \phi_q$ with undergoing semiclasical evolution   $\delta \phi_c$.  For a given field, the sign of  $\delta \phi_q$ is random, while the semiclassical motion always points in the downhill direction. The {\em average} field thus has a stochastic fluctuation  $ \delta \phi_q/\sqrt{N}$, since this is the mean of $N$ signed, random variables.  Consequently, the collective motion of the ensemble of fields remains deterministic, and the density of the universe will decrease monotonically with time unless $\delta \phi_q \gsim \sqrt{N}  \delta \phi_c$. In this case the density fluctuation is boosted to  order unity, and the average density can increase inside a given Hubble patch, so inflation  is not just stochastic but eternal.  Since $H \propto \rho^{1/2}$, we are in the intermediate range where inflation is stochastic but not eternal,  $\delta \phi_q /\delta \phi_c $ diminishes with time. In this case the stochastic motion of the {\em individual\/} fields will eventually become subdominant and for any reasonable value of $N$ the inflationary dynamics will be well-described by the semi-classical motion alone. 

From a practical perspective, a period of stochastic inflation in a simple multi-field model need not modify the observable properties of the universe. Certainly in the case described above, the stochastic motion would cease well before $P_{\cal R}^{1/2} \sim 10^{-5}$ unless $N$ is very large. However, since the stochastic motion necessarily increases the variance in the individual field values this phase may have an impact on the likely initial spread in the values the $\phi_I$, which can have an impact on the inflationary observables in $N$-flation. However, there is no clear expectation for the likely initial values of the $\phi_I$ and without this the impact of any stochastic evolution cannot be evaluated. 

On the other hand, if the individual potential terms do not all have a single well-defined minimum, any phase where one of more of the individual fields moves stochastically could have a substantial impact on the inflationary phenomenology.    For instance, in the case of $N$-flation, the $N$ fields are actually axions and thus have periodic potentials. The individual $m^2 \phi_I^2$ terms arise from assuming that each field is close to it minimum, when measured relative to the scale on which the underlying potential is periodic, and then Taylor expanding.  However, if we imagine an initial state when the $N$ fields (or even some subset of them) are close to the maxima of their cosine potentials two effects will occur. The first is that these fields will have a  small $\delta\phi_c$, since the corresponding $V_I'$ will be very small in the vicinity of an extremum, making it more likely that these fields move stochastically even if other fields are dominated by the semiclassical rolling. Secondly, with fields moving stochastically near the maxima  of these potentials, in individual Hubble domains in which these fields do evolve away from the peaks the fields will then roll towards different minima. In {\em eternal\/} inflation each patch in which the stochastic motion ceases would be identified as a ``pocket  universe'' and this scenario thus produces pockets with many different vacua, depending on the symmetry breaking pattern of the individual axions. However,  if enough fields are rolling semi-classically, $H$ is strictly decreasing, and inflation is not future-eternal. In this case we can end inflation globally, and count the number and type of distinct domains created during the stochastic phase.  Consequently we now have a toy model that initially resembles an eternally inflating universe, but in which there is a natural late-time cutoff which removes the infinities which otherwise can prevent one calculating the relative creation rates for different types of pocket.  This system may thus prove useful for testing different ``measures'' for eternal inflation  \cite{Linde:1993xx,Guth:2000ka,Garriga:2005av,Easther:2005wi, Bousso:2006ev, Aguirre:2006ak,Easther:2007sz}, and we will examine this issue in a separate publication. Conversely, once inflation has ended, these different domains will eventually merge and the late-time  universe (in the absence of a cosmological constant term) will be composed of domains with different vacua, separated by domain walls. 

%-------------------------------------------------------------------------------------------------------------
%-------------------------------------------------------------------------------------------------------------
\section{One-loop Quantum Corrections} \label{sect:loop}
%-------------------------------------------------------------------------------------------------------------
%-------------------------------------------------------------------------------------------------------------

In the previous section, we considered  stochastic field evolution in cosmological scenarios with many degrees of freedom. We now assume the field evolution is well-described by the semi-classical equations of motion, and compute loop corrections to the perturbation spectrum.  Scalar loop corrections to the  inflationary spectrum in single field models have been widely discussed  \cite{Weinberg:2005vy,Seery:2005gb,Weinberg:2006ac,Sloth:2006nu,Sloth:2006az,Prokopec:2008gw}.  Likewise, the correction  from graviton loops was computed in \cite{Dimastrogiovanni:2008af}.\footnote{ In this work we are primarily interested with the scaling behavior of the loops as one increases the number of fields. Since there are only 2 graviton modes, loops involving gravitons cannot scale with the number of fields $N$ (even though individually they are of the same order), and hence we neglect them. }  Weinberg \cite{Weinberg:2005vy} looks at loop corrections to a single inflaton in the presence of many massless spectator fields, and we now generalize this result to models with $N$ participator fields.  For clarity, we present our results in this section, and discuss technical aspects of the calculation in  Appendix \ref{app:1loop}.

We previously made the distinction between single field inflation with $N$ spectator fields, and assisted scenarios with $N$ participator fields, of which $N$-flation is the most interesting example.  The field dynamics differs between these cases, since only the participator fields have non-zero VEV and contribute to the vacuum energy that drives inflation.  However, both classes of field  can contribute loop corrections to the 2-point function or perturbation spectrum -- and we will see that the {\em forms\/} of these contributions are different.  The immediate concern  is that loop corrections result in a new ``species problem'', as pointed out in  \cite{Veneziano:2001ah,Dimopoulos:2005ac,ArkaniHamed:2005yv}.  Generically, if we insist that    quantum corrections to the Planck scale are small, we need   $\sim N(\Lambda_{UV}/M_p)^2$ to be small, where $\Lambda_{UV}$ is some UV cut-off scale. Having computed the loop corrections to the 2-point \emph{correlation} function, we can then determine whether these also  provide an  constraint on $N$, as a function of $\Lambda_{UV}/M_p$. Surprisingly, with $N$ participator fields, there is no bound on $N$. On the other hand, with $N$ spectator fields, we obtain slightly weaker bounds on $N$ as compared to the standard arguments.  In what follows we look at the two limiting cases, firstly $N$-spectator fields with a single inflaton and then $N$ participator fields.  One could easily generalize our result to the case where one had a mix of both spectator and participator fields.

%-------------------------------------------------------------------------------------------------------------
\subsection{One-loop corrections of $N$ Spectator fields}
%-------------------------------------------------------------------------------------------------------------

Consider inflation driven by a single field field $\phi$, with $N$ spectator fields, $\sigma_I$ where $I$ runs from $1$ to $N$,
\begin{equation}
S = \int dx^4 \sqrt{g}\left[\frac{M_p^2}{2}R - \frac{1}{2}\left(\partial \phi\right)^2 + V(\phi) - \sum_I \frac{1}{2}\left(\partial \sigma_I\right)^2\right]. \label{eqn:weinbergcaseaction}
\end{equation}
In the discussion below, repeated indices are not summed over unless explicitly specified. Unlike the $N$-flation case, the   spectator fields remain invariant under the shift $\sigma_I \rightarrow \sigma_I + \delta_I$. Any initial kinetic energy possessed by these fields decays away rapidly, since $\rho_{i} \propto (\dot{\sigma}_i)^2 \propto a^{-6}$.  Thus the only contribution to the  background energy density is the inflaton potential,  $V(\phi)$. This situation has been discussed by 
Weinberg \cite{Weinberg:2005vy}, who found that the one-loop two-vertex quantum corrections modify the power spectrum equation (\ref{eqn:zeroPS}) by a term of order $(H/M_p)^4  \ln k$ per field. Thus, for $N$ $\sigma_I$ fields\footnote{The numerical factor in  \cite{Weinberg:2005vy} differs from that of equation (\ref{eqn:weinberg1loop}). As we explain in Appendix \ref{app:1loop}, there is an extra  contribution, related to the contour in the time integral needed to pick up the ``in'' vacuum. We thank Steven Weinberg for a useful discussion of this point.}  the first order correction to the power spectrum is 
\begin{equation}
P_k^{(1)} = \frac{1}{4(2\pi)^3}N\frac{\pi}{6} \frac{H^4}{M_p^4} \epsilon\ln k. \label{eqn:weinberg1loop}
\end{equation}
The one-loop corrected power spectrum is then
\begin{equation}
P_k\rightarrow \frac{1}{4(2\pi)^3}\frac{1}{\epsilon}\frac{H^2}{M_p^2}\left(1+(c_1+N \frac{\pi}{6}\epsilon)\frac{H^2}{M_p^2}  \ln k \right),
\end{equation}
where $c_1 = -2\pi/3$ is the one-vertex self-correction term which we compute in the appendices. What about the one-loop one-vertex loops of the $\sigma$ fields around the inflaton? As we explained in detail in Appendix \ref{app:1loop} and later in Section \ref{sect:participator}, there is no non-scale-free $\sigma_I$ one-vertex correction to the $\phi$ propagator, and hence Weinberg's computation is complete modulo the inflaton self-correction. 

Requiring that the one-loop corrections do not dominate the ``tree-level'' two-point correlation, we obtain the bound
\begin{equation}
N<\frac{M_p^2}{H^2}\frac{1}{\epsilon}.
\end{equation} 
For successful slow roll inflation $\epsilon \ll 1$, so this bound is necessarily weaker than that obtained from gradient energy considerations. Interestingly, the ``tree-level'' power spectrum for the tensor modes for single scalar field inflation is 
\begin{equation}
P_{\rm gw} = \frac{H^2}{M_p^2}.
\end{equation}
Hence any measurement of $P_{gw}$ would effectively put an \emph{observational} upper bound on $N$,
\begin{equation}
N < 1/P_{\rm gw}.
\end{equation}

  %-------------------------------------------------------------------------------------------------------------
\subsection{One-Loop corrections with $N$ Participator Fields} \label{sect:participator}
%-------------------------------------------------------------------------------------------------------------

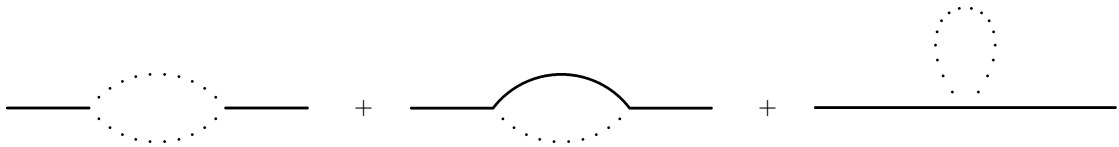
\begin{figure}
\begin{fmffile}{fig1}
\unitlength =1mm
\begin{eqnarray}
\parbox{40mm}{
\begin{fmfgraph*}(40,25)
\fmfleft{in}
\fmfright{out}
\fmf{plain}{in,v1}
\fmf{plain}{v2,out}
\fmf{dots,label=$J$,left=0.5,tension=0.3}{v2,v1} 
\fmf{dots,label=$J$,left=0.5,tension=0.3}{v1,v2}
\fmflabel{$I$}{in}
\fmflabel{$I$}{out}
\end{fmfgraph*}}
~~~~~+~~~~
\parbox{40mm}{
\begin{fmfgraph*}(40,25)
\fmfleft{in}
\fmfright{out}
\fmf{plain}{in,v1}
\fmf{plain}{v2,out}
\fmf{dots,label=$J$,left=0.5,tension=0.3}{v2,v1} 
\fmf{plain,label=$I$,left=0.5,tension=0.3}{v1,v2}
\fmflabel{$I$}{in}
\fmflabel{$I$}{out}
\end{fmfgraph*}}
~~~~~+~~~~
\parbox{40mm}{
\begin{fmfgraph*}(40,25)
\fmfleft{in}
\fmfright{out}
\fmf{plain}{in,v1,out}
\fmf{dots, label=$J$, tension=1}{v1,v1}
\fmflabel{$I$}{in}
\fmflabel{$I$}{out}
\end{fmfgraph*}}
\nonumber
\end{eqnarray}
\end{fmffile} 
\caption{Feynman diagrams for the three- and four-point one-loop correction for the $I$ field correlator. With $N$ participator fields, the first two terms apparently scale as $N$, due to the summation of $J$,   but each loop is suppressed by $\epsilon/N$, arising from the extra coupling. At leading order in slow roll, the third term only has contributions from the self-interaction $I=J$ term, so the corrections do not blow up as $N$ becomes large. Here we have denoted $I$ fields by a solid lines while dotted lines represent $J$ fields.} \label{fig:loopdiagrams}
\end{figure}

We now turn to the case with $N$ participator fields, N-flation, which is based on the action of equation (\ref{eqn:Nflationaction}), and we  perturb each individual field,
\begin{equation}
\phi_I \rightarrow \bar{\phi_I} + Q_I,
\end{equation}
where $\bar{\phi_I}$ is the homogeneous background solution and $Q_I$ is the perturbation. In the discussion below, upper case Roman letters $I,J,...$ label the fields, and we have a flat target space $X^I = X_I$ with the summation convention
\begin{equation}
X^I Y_I = \sum_I X_I Y_I \, .
\end{equation}
The power spectrum produced by $N$ uncoupled inflating fields is \cite{Sasaki:1995aw}
\begin{equation}
P_k = \frac{1}{N^2} \sum_I \left(\frac{H}{\dot{\phi}_I}\right)^2\langle Q_I Q_I  \rangle, \label{eqn:nfieldPS}
\end{equation}
The $1/N^2$ factor in front reminds us that the total power spectrum is not simply a sum of the individual power spectra of the fields; the power spectrum is itself the expectation value of a set of random variables.  In general, 
\begin{equation}
\langle Q_I Q_I \rangle = \langle Q_I Q_I \rangle_0 + \langle Q_I Q_I \rangle_1 + \langle Q_I Q_I \rangle_2...,
\end{equation}
where the numerical subscript denotes the order of the expansion, or the number of vertices.  The uncorrected power spectrum is simply
\begin{equation}
\langle Q_I Q_I \rangle_0 = \frac{1}{2 (2 \pi)^3}H^2, \label{eqn:noloopifield}
\end{equation}
leading to the primordial power spectrum,
  \begin{equation}
  P_k = \frac{1}{4(2\pi)^3} \frac{H^2}{M_p^2 N \epsilon_I}. \label{eqn:zeroPS}
  \end{equation}
Our goal is to compute the one-loop corrections to the power spectrum for each field $\langle Q_I Q_I \rangle_{\rm 1-loop} $ generated by all $N$ fields. 
To compute the one-loop correction, we use   (\ref{eqn:ininformula}), with the appropriate interaction Hamiltonian $H_{\rm int}$ for each field.\footnote{ All the fields below are the taken to be Heisenberg fields unless otherwise noted. 
} The three-point interaction Hamiltonian is, to leading order in slow roll \cite{Seery:2005gb},
  \begin{equation}
  H_{\rm int}^{(3)}(t) = \int d^3 x \left[\frac{a^3}{4H} \sum_{I,J}\dot{\phi}_I Q_I \dot{Q}_J \dot{Q}_J + \frac{a^3}{2H} \sum_{I,J}\dot{\phi}_I \partial^{-2} \dot{Q}_I \dot{Q}_J \partial^2 Q_J\right]. \label{eqn:NloopHI}
  \end{equation}
  
Application of equation (\ref{eqn:2pt2}) leads to three sets of diagrams generated by the two interaction terms and their cross-interaction, the 1st two diagrams of Fig. \ref{fig:loopdiagrams} where each vertex corresponds to an interaction via one of the terms above. We also need the four-point action, equation (\ref{eqn:big4pt}) derived in Appendix \ref{App:4ptderivation},  
  \begin{eqnarray}
  H_{\rm int}^{(4)}(t) & =&  \int d^3 x a^{3}\left[ \frac{1}{4Ha^{2}}\sum_{I,J} \partial_i Q_J\partial_iQ_J\partial^{-2}  (\partial_j \dot{Q}_I\partial_jQ_I + \dot{Q}_I\partial^2 Q_I) \right. \nonumber \\
  && + \frac{1}{4H} \sum_{J,I} \dot{Q}_J \dot{Q}_J \partial^{-2}(\partial_i \dot{Q}_I \partial_j Q_I + \dot{Q}_I \partial^2Q_I) \nonumber \\
  && +\frac{3}{4}\sum_{I,J}\partial^{-2}  (\partial_j \dot{Q}_J \partial_j Q_J + \dot{Q}_J\partial^2 Q_J)\partial^{-2} (\partial_j \dot{Q}_I \partial_j Q_I + \dot{Q}_I \partial^2 Q_I) \nonumber \\
&&\left. +\frac{1}{4}\beta_{2,j}\partial^2 \beta_{2,j}+\sum_{I}\dot{Q}_I\partial_i Q_I\beta_{2,i}\right],
\end{eqnarray}
where $\beta_{2}^j$ is given in equation (\ref{eqn:beta2j}) and  repeated lower roman indices $i,j,\cdots$  are summed using the Euclidean metric; $a_{i}b_{i} = \sum_{i,j}\delta_{ij}a_{i}b_{j}$. At first order this interaction generates the third diagram of Fig. \ref{fig:loopdiagrams} via Eq. (\ref{eqn:1pt}). For simplicity, we ignore the self-interaction term of the two-vertex term, as we expect that its correction to be of the same order as the other $N-1$ corrections \cite{Seery:2007wf,Seery:2007we}.  

Note that in deriving the above  expression we have made use of  the substitution $\mathcal{H}_{int} =  -{\cal L}_{int}$, where ${\cal L}_{int}$ is the Lagrangian density computed to the appropriate order in the perturbation. This substitution is trivial with non-derivative interactions, but it is more subtle in the presence of derivative interactions. However, in Appendix \ref{app:canquant}  we show that we are effectively ignoring a correction that is at most of order ${\cal O}(\epsilon)$,
\begin{equation}
\mathcal{H}_{\rm int} = -{\cal L}_{\rm int} + {\cal O}(\epsilon),
\end{equation}
which can be discarded to leading order in slow-roll.

Each of the field $Q^I$ can be expanded in Fourier modes and their respective creation/anihilation operators,
\begin{equation}
Q^I(\bdv{x},t) =\int d^3 k\,e^{i\bdv{k}\cdot\bdv{x}} \left[a_I(\bdv{k})\U{k}{I}(t)+a_I^{\dagger}(\bdv{-k})\Ud{k}{I}(t)\right], \label{eqn:Qexpansion}
\end{equation}
where the functions $U^{I}_{k}(t)$ are solutions of the equations of motion obtain from varying the second order action, Eq. (\ref{eqn:quadraticaction}), with respect to the field $Q^{I}$. The $a^{I}({\bf k})$ satisfy the usual commutation relations,
\begin{equation}
[a_I(\bdv{k}),a_J^{\dagger}(\bdv{k}')]=\delta(\bdv{k}-\bdv{k}')\delta^I_J.\label{eqn:Qcommutator}
\end{equation}
With these ingredients, we make the final simplifying assumption that the fields are identical, which means that we simply need to compute the one-loop correction from one of the $J$ terms in equation (\ref{eqn:NloopHI}) and then multiply them by $N-1$ to get the total correction per field $I$. We leave the  details of this computation to Appendix \ref{app:1loop}. 

It turns out that the only the two-vertex diagrams and the one-vertex \emph{self-interaction} diagram contribute the physical logs while \emph{non-self-interaction} diagrams of the one-vertex loops contribute polynomial ultra-violet divergences which we assume are absorbed by renormalization.   This is a consequence of the symmetry of the original action where the potentials are uncoupled and the kinetic terms are $SO(N)$ symmetric.  This can be understood as follows.  The diagrams in Fig.(\ref{fig:momentumstructure})   have metric ``graviton'' propagators and loops hidden inside them -- the higher order interactions that appear perturbatively are mediated by gravity. However, we have chosen a gauge where the metric perturbation vanishes. If we choose a different gauge (Newtonian gauge \cite{Mukhanov:1990me} for example) these interactions are manifest, and each one-vertex diagram receives contribution from terms  like
\begin{eqnarray}
{}\nonumber \\
{}\nonumber \\
{}\nonumber \\
\parbox{40mm}{
\unitlength =1mm
\begin{fmffile}{fig2a}
\begin{fmfgraph*}(40,15)
\fmfleft{in,in1}
\fmfright{out,out1}
\fmf{phantom}{in1,v2,out1}
\fmf{plain}{in,v1}
\fmf{plain}{v1,out}
\fmf{wiggly,pull=10,left=0.1,tension=0.3}{v1,v2} 
\fmf{plain,label=$J$}{v2,v2}
\fmflabel{$I$}{in}
\fmflabel{$I$}{out}
\end{fmfgraph*}
\end{fmffile}}
~+~~~~~
\parbox{40mm}{
\unitlength =0.7mm
\begin{fmffile}{fig2b}
\begin{fmfgraph*}(40,15)
\fmfleft{in}
\fmfright{out}
\fmf{plain}{in,v1}
\fmf{plain}{v2,out}
\fmf{wiggly,left=0.5,tension=0.3}{v1,v2} 
\fmf{plain,left=0.5,label=$J$,tension=0.3}{v2,v1}
\fmflabel{$I$}{in}
\fmflabel{$I$}{out}
\end{fmfgraph*}
\end{fmffile}}
~+~\cdots\\
{}\nonumber
\end{eqnarray}
where the wiggly lines represent the graviton lines, the solid lines are the field perturbation lines and the ellipses denote terms that are higher order in metric perturbation or the slow roll parameters. The first ``balloon'' diagram diverges polynomially; the loop that runs around itself does not possess a scale since it is not directly connected to an external leg.  Both vertices of the second term must obey the symmetry of the original action, which forces $I=J$. It is clear from this perspective why only the self interaction one-vertex loops can contribute to physical obervables.\footnote{We thank Dan Kabat for a very useful discussion on this point.}

After regularizing the results and assuming that we can always find suitable counterterms to cancel the gravitational backreaction and UV divergences from loops independent of external momenta, the correction of $J$ loops to the $I$ propagator is given by equations (\ref{eqn:term1result}) and (\ref{eqn:4ptselfcorrection}). The total correction is thus
\begin{equation}
\langle Q_I Q_I \rangle_{\rm 1-loop} =  \frac{1}{2(2\pi)^3}\frac{H^4}{M_p^4}\left[c_1 + N c_2 \frac{\dot{\phi}_I^2}{M_p^2 H^2}\right] \ln k, \label{eqn:1loopifield}
\end{equation}
where $c_1 = -2\pi/3$ and  $c_2=(2017/240)\pi$, which  arise from   the   one-vertex self-interaction and the two-vertex loops respectively. We have set $N-1 \rightarrow N$, since $N\gg1$, so the two-vertex self-interaction is not important.     During $N$-flation,   each participator field is corrected by its $N-1$ counterparts in the two-vertex loop, and by itself in the one-vertex loop. However, in equation (\ref{eqn:1loopifield}), the two-vertex corrections are suppressed by the {\em  individual\/} ``slow-roll'' parameter,
  \begin{equation}
  \epsilon_I \equiv  \frac{1}{2}\frac{\dot{\phi}_I^2}{(H M_p)^2}.
  \end{equation}
Since $\dot{\phi}_I^2$ is the velocity of a single field, while $H^2$ is related to the overall density, this quantity is reduced relative to its value in the single field case by a factor of $N$. We can think of $\epsilon_I$ as the square of the coupling strength of each three-point vertex in the interaction Hamiltonian equation (\ref{eqn:NloopHI}).
 The total correction to the power spectrum is obtained by inserting in equations (\ref{eqn:noloopifield}) and (\ref{eqn:1loopifield}) into (\ref{eqn:nfieldPS}),
  \begin{equation}
  P_k = \frac{1}{4N^2(2\pi)^3}\sum_{I} \frac{1}{\epsilon_I}\frac{H^2}{M_p^2}\left[1+(c_1+ 2c_2N \epsilon_I)\frac{H^2}{M_p^2}\ln k\right]. \label{eqn:finalNPS}
  \end{equation}
Now for slow roll,
  \begin{equation}
  \epsilon \equiv -\frac{\dot{H}}{H^2} = N \frac{1}{2}\frac{\dot{\phi}_I^2}{H^2 M_p^2} =N\epsilon_I \ll 1, \label{eqn:coherentepsilon}
  \end{equation}
  and hence the power spectrum can be rewritten as
  \begin{equation}
P_k = \frac{1}{4(2\pi)^3}\frac{H^2}{M_p^2}\frac{1}{\epsilon}\left[1+(c_1+ 2c_2 \epsilon) \frac{H^2}{M_p^2} \ln k \right],
  \end{equation}
  which is equivalent to the power spectrum of a single scalar field inflation $\varphi$ with its one-loop self-correction \cite{Seery:2007we}.
 
The lack of any  $N$-dependence in this bound is somewhat surprising. In fact, we will now show that this non-appearance is true to $\emph{all orders}$ in loops, provided slow roll can be assumed. The pair-wise $SO(N)$ symmetry $X^I X_I$ in the kinetic term of the action is preserved when we expand the action to higher orders.\footnote{The   flat  target space  i.e. $G^{IJ}\nabla_{\mu} \phi_I \nabla^{\mu} \phi_J$ with $G^{IJ}$ diagonal is crucial here. The pair-wise symmetry will not hold if $G^{IJ}$ is not diagonal, and we cannot write down the theory as a single equivalent coherent scalar field. Another way of seeing this that one can always redefine the fields so that the target space is flat at the cost of generating couplings, both direct and gravitational, in the potential.}  Consequently, the index structure of  terms in $H_{\rm int}$ is always fully pair-wise e.g. $\dot{\phi}^I Q_I Q^J Q_J$, $Q^I Q_I Q^J Q_J$ or $\dot{\phi}^I \dot{\phi}^J Q_I Q_J Q^K Q_K$ etc. since the interactions must preserve this same symmetry.  When computing the one-loop diagram, we have to contract \emph{through} the loop to obtain the physical log contribution, in the sense that we have to contract the external lines with \emph{at least one} of the interacting fields in the loop. We have already argued above and in the Appendix \ref{app:1loop} that there can be no cross-coupling between $I$ and $J$ fields in the four-point interaction, at least at lowest order in slow-roll. Interaction terms like $\dot{\phi}_J \dot{\phi}_K \partial^{-2}(Q^I Q^I) Q^J Q^K$ do exist and can contribute a log divergence, however, these are higher order in slow-roll thanks to the extra $\dot{\phi}$ terms and so their contribution to the one-loop correction will be $\epsilon/N$ suppressed.

Let us now turn to the diagrams generated by the three-point interaction and the even vertex loops. For our coherent field argument to be true, the \emph{leading} $2n$-vertex correction must be of the order  $(N\epsilon_I)^n \equiv \epsilon ^n$. We can now count the number of diagrams to find the factors of $N$, remembering that  each time a $\dot{\phi}^I \propto \sqrt{\epsilon^I}$ coupling appears, we get a factor of $\sqrt{1/N}$ from the coherent field relation equation (\ref{eqn:coherentepsilon}).  Consider the simplest diagram which we have calculated in the Appendix: a two-vertex loop with identical interaction term $\dot{\phi}^I Q_I Q^J Q_J$ at both vertices. We now want to count the factors of $N$ and $\epsilon_I$ for the correction to the $Q^I$ propagator. Since we have to contract \emph{through} the loop, we can only contract the $Q^J$ of one interaction to the $Q_J$ of the other interaction (else we will form a disconnected diagram if we contract the $J$ fields at the same space point). Hence, by counting sums, we get a factor of $N$ from the $J$ contraction and a factor of $\epsilon_I$ from the couplings, for a total correction of $N\epsilon_I = \epsilon$, equivalent to the correction for a single coherent field  as we have shown in the above.

Next, consider a more complicated interaction term $\dot{\phi}^I \dot{\phi}^J \dot{\phi}^K Q_I Q_J Q_K$. Since the fields and couplings have to appear pair-wise, there is no ${\cal{O}}(\dot{\phi}^2)$ interaction. There are two ways to contract through the loop, i.e. via $Q^J$ and $Q^K$, so we get two factors of $N$. However, the three factors of $\dot{\phi}^2$ give us the coupling term $\epsilon_I \epsilon_J \epsilon_K$ , and  equation (\ref{eqn:coherentepsilon}) yields the correction term $N^2\epsilon_i^3 = \epsilon^3/N$, which is $1/N$ suppressed compared to the previous case, and hence contributes even less. Adding loops will add more factors of $\epsilon_I$ into the interaction and thus provide further $1/N$ suppressions -- it is easy to see that the extra factors of $N$ coming from the extra loops will never scale more than the suppression that comes from $\epsilon_I = \epsilon/N$ .   The point here is clear: since the field labels have to appear pair-wise, we cannot have three-point interactions like $\dot{\phi}_I  Q^I Q^J Q^K$ that could have given us an $\sim N \epsilon$ correction that has no coherent field analog. This argument does not depend on the number of loops, since it depends solely on the interaction terms. 
 
In fact, we can write the $N$ fields in terms of a single scalar field model, where the inflaton is the radial field.  Assume that each individual field's potential is  $(1/2)m_I^2 \phi_I^2$ and that they have identical masses $m_I = m$. We can then rewrite the fields in polar coordinates, $\psi^2 = \sum_I \phi_I^2$, in which case the Lagrangian becomes  \cite{Dimopoulos:2005ac} 
\begin{equation}
\mathcal{L} =\frac12 (\partial \psi)^2 - \frac12 m^2 \psi^2  +\frac12 \psi^2 (\partial \Omega)^2,  \label{eqn:coherentaction}
\end{equation}
where $\langle \psi^2 \rangle \sim (N\epsilon_I)^{-1}(H/M_p)^2$. The angular terms  thus have large values and damp out quickly, dropping out of the inflationary dynamics.  Consequently, the set of $N$ coherent fields can be recast as theory with a single scalar field, and we are thus unable to derive a bound on $N$.

One might worry that the direct coupling term $\psi^2(\partial\omega_I)^2$ with $(\partial \Omega)^2 = \sum (\partial\Omega_I)^2$, will generate $N-1$ loop corrections that are not scale-free. However, we will sketch below that this interaction can, at the most, generate scale-free loops and hence is harmless.
Consider the perturbed fields
\begin{eqnarray}
\psi& \rightarrow& \bar{\psi} + Q, \\
\Omega_I &\rightarrow & \bar{\Omega}_I + \omega_I.
\end{eqnarray}
Using the symmetry of the interaction, one can show that the leading three-vertex interactions are $ \bar{\Omega}_I QQ \omega^I$ and $\bar{\psi} Q \omega_I \omega^I$ while the leading four-vertex interaction is $QQ \omega_I \omega^I$, each with various permutations of derivatives. The first three-vertex term can at the most generate a single term, since the attractor solution drives the background angle fields to a fixed trajectory and hence we can always pick $\Omega_I = (1,0,0,...,0)$. For the latter three-vertex term, we have to contract through 4 instances of $\omega_I$, but it can be shown that the propagator for $\omega_i \propto a^{-3}$, and hence the loops are quickly redshifted away. Meanwhile it is clear from our discussion above that the four-point interaction can at most generate  scale-free terms. For example, the lowest order diagram is the one-loop one-vertex diagram; since the external legs are $Q$'s and the $\omega_I$ can only contract with itself this is scale free.\footnote{The analog in the $\phi_I$ field picture will be the ``balloon'' diagram.}

\section{Discussion}  \label{sect:conclusions}

In this paper we explore modifications to the dynamics of inflationary models in the presence of $N$ light scalar degrees of freedom. We make a distinction between spectator fields, which do not contribute to the background energy density, and participator fields, whose potential terms contribute to the inflationary background.  As summarized in the introduction, a number of very general arguments place finite upper bounds on $N$, which typically take the form $N\ll M_p^2/H^2$.  
We first consider the dynamics of stochastic inflation in the presence of large number of light fields, and show that when $N$ is large there is a distinction between stochastic and eternal inflation which does not apply in the single field case. In particular, with a large number of participator fields we show that there is a regime where the individual field motion is dominated by quantum fluctuations and thus stochastic, while the overall evolution of the universe is deterministic.  Moreover, if the stochastic fields have symmetry breaking potentials, then one can create a large number of apparent ``pocket'' universes, while retaining the ability to end inflation globally, and controlling the divergences characteristic of scenarios in which inflation is genuinely eternal.

Secondly, we explore loop corrections to the 2-point correlation function that provides the inflationary perturbation spectrum. Since any light field can run round a loop, these will typically scale with $N$. We analyze two subcases -- a single inflaton with $N$ spectator fields, and $N$ participator fields. In the former case, we find an explicit bound, but one which is weaker (by one over a factor that must be small during slow roll) than the simple form described above, namely
\begin{equation}
N\lsim \frac{M_p^2}{H^2}\frac{1}{\epsilon}.
\end{equation}
On the other hand, with $N$ participator fields ($N$-flation type scenarios), the loop correction is small and independent of $N$.  We can understand this result by recasting the action in terms of a composite single scalar field.  Finally, in the course of this work, we have had need to look closely at the computation of loop corrections in the in-in formalism.  These calculations raise a number of subtle issues, and we give details of our approach in the Appendices.

One might  whether bounds on $N$ actually rule out otherwise realistic fundamental theories. Within string theory, Vafa \cite{Vafa:2005ui} argued that $T$ and $S$ dualities ensure that the volume of scalar moduli space is finite, and that  there is a strict upper bound on the number of matter fields.   We are not aware of explicit stringy constructions which would saturate the large $N$ bounds described above  in any reasonable compactification of string theory and with a finite non-zero Newton constant. 

As we have noted, many independent arguments put limits on the allowed value of $N$.      For example, reference \cite{Watanabe:2007tf} notes that if $N$ is too large, the inflaton can decay into a large number of  species during reheating, which may cause phenomenological problems in the later universe, while  \cite{Leblond:2008gg} suggests that validity of the perturbative expansion itself provides a constraint on $N$.    We have chosen to work in the simplest models of multi-field inflation, where the fields have uncoupled potentials. One can  multifield hybrid inflation \cite{Linde:1993cn} with large number of coupled fields, and it would be interesting to ask if such models which are not ruled out by radiative corrections to their potentials but by the loop corrections to their power spectrum.   Also, while higher correlation functions  themselves do not seem to impose any bound on $N$ \cite{Battefeld:2006sz}, one could check whether this was also true of their quantum corrections.

\section{Acknowledgments}
We thank  Nicola Bartolo, Emanuela Dimastrogiovanni, Bei Lok Hu, Daniel Kabat, Eiichiro Komatsu,  Louis Leblond, Liam McAllister, Alberto Nicolis, David Seery, Sarah Shandera and Steven Weinberg for a number of useful conversations. We are particularly grateful to  Walter Goldberger for a number of extremely helpful suggestions.  
 RE is supported in part by the United States Department of Energy, grant DE-FG02-92ER-40704 and by an NSF Career Award PHY-0747868.  This research was supported by grant RFP1-06-17 from The Foundational Questions Institute (fqxi.org) 

%-------------------------------------------------------------------------------------------------------------
%-------------------------------------------------------------------------------------------------------------
\appendix
%-------------------------------------------------------------------------------------------------------------
%-------------------------------------------------------------------------------------------------------------
  \section{The ``in-in'' formalism and one loop corrections to two-point correlation functions for multifield models} \label{app:1loop}

In this Appendix, we review the canonical quantization approach to the Schwinger-Keldysh ``in-in'' formalism  \cite{Schwinger:1961,Keldysh:1964ud} and use it to compute the one-loop correction\footnote{See also \cite{Riotto:2008mv} for a discussion of beyond one-loop effects.} to the $I$-th field two-point correlation function. There are two types of loops: a two vertex loop of the type considered by Weinberg \cite{Weinberg:2005vy}, and a one vertex loop of the type considered by Seery \cite{Seery:2007we} which we calculate below.

The Schwinger-Keldysh ``in-in'' formalism \cite{Schwinger:1961,Keldysh:1964ud}, was first applied to cosmology by Jordan \cite{Jordan:1986ug} and Calzetta and Hu \cite{Calzetta:1986ey}. This formalism was reintroduced into the computation of cosmological correlations by Maldacena \cite{Maldacena:2002vr} and extended beyond tree-level by Weinberg \cite{Weinberg:2005vy}. We follow Weinberg's notation and methods, with some comments on its relationship with the functional method of \cite{Jordan:1986ug,Calzetta:1986ey}.  
 The correlation that we want to compute is
\begin{equation} \label{eqn:ininformula}
\langle W(t)\rangle =\left\langle \left(T e^{-i\int_{-\infty}^{t}H_{\rm int}(t) dt}\right)^{\dagger}  ~  W(t)~ \left(Te^{-i\int_{-\infty}^{t} H_{\rm int}(t) dt}\right)\right\rangle,
\end{equation}
where $W(t)$ is some product of fields,  $H_{int}$ is the interaction Hamiltonian, both which are constructed out of Heisenberg (free) fields and $T$ is the time-ordering symbol. The expectation is taken over the true ``in'' vacuum.

One can use the Dyson expansion for the time evolution operator, together with the Baker-Campbell-Hausdorf formula to express equation (\ref{eqn:ininformula}) in what may appear to be a more convenient form \cite{Weinberg:2005vy};
\begin{eqnarray}\label{eqn:Wein2}
\langle W(t) \rangle = \sum_{N = 0}^{\infty}i^{N}\int_{-\infty}^{t}dt_{N}\int_{-\infty}^{t_{N}}dt_{N-1}...\int_{-\infty}^{t_{2}}dt_{1}\left\langle \left[H_{\rm int}(t_{1}), \left[H_{\rm int}(t_{2}), ... \left[H_{\rm int}(t_{N}), W(t)\right]...\right]\right]\right\rangle.
\end{eqnarray}
While technically equivalent, for these computations this form proves problematic. To perform calculations one assumes that at very early times the vacuum state is the bare vacuum, that is, that the interactions disappear. Operationally, one implements this by deforming the time contour in equation (\ref{eqn:ininformula}) off the real axis into the lower half plane to include a small amount of evolution in imaginary time, killing off the interactions in the far past ($-\infty\rightarrow-\infty(1+i\epsilon)$). Unfortunately, in equation (\ref{eqn:Wein2}) this scheme can not be easily implemented.  Contours entering from the right and the left side of equation (\ref{eqn:ininformula}) are treated identically when deriving equation (\ref{eqn:Wein2}). However, once the vacuum prescription has been specified these contours are in fact complex conjugates of each other, and can no longer be freely interchanged.\footnote{Note that this issue is not manifest at tree level, and the conclusions of  \cite{Weinberg:2005vy,Weinberg:2006ac} are robust, other than with respect to the coefficient of the $N$ spectator loop correction,  as discussed in Section \ref{sect:loop}. Musso \cite{Musso:2006pt} has developed a diagrammatic formalism for correlation functions in the ``in-in'' formalism.  However, since this is based primarily on equation (\ref{eqn:Wein2}), one would need to check it carefully before employing it for an explicit calculation.}
 In the rest of this work, we work directly with equation (\ref{eqn:ininformula}), expanding it to the desired order. 

At first order we have, using hermiticity,
\begin{equation}\label{eqn:1pt}
\langle Q^{I}(t)Q^{I}(t) \rangle_{1} = -2\Im \int_{-\infty_{-}}^{t} dt_1 \langle H^{(4)}_{\rm int}(t_1) Q^I(t)Q^I(t)\rangle,
\end{equation}
where we have introduced the shorthand $\infty_{\pm}\equiv \infty(1\pm i\epsilon)$. Notice that equation (\ref{eqn:1pt}) is manifestly real. This reality is not surprising: we are computing correlation functions and not transition amplitudes as noted in \cite{Calzetta:1986ey,Jordan:1986ug}. At second order we have, again using hermiticity,
\begin{eqnarray}\nonumber\label{eqn:2pt2}
\langle Q^I(t) Q^J(t)\rangle_{2} & = & - 2\Re \int^{t}_{-\infty_{-}} dt_2 \int^{t_2}_{-\infty_{-}} dt_1 \langle H^{(3)}_{\rm int}(t_1)H^{(3)}_{\rm int}(t_2)Q^I(t) Q^J(t)\rangle \\ && +\int_{-\infty_{-}}^{t}dt_{1}\int_{-\infty_{+}}^{t}dt_{2}\langle H^{(3)}_{\rm int}(t_1)Q^I(t)Q^I(t)H^{(3)}_{\rm int}(t_2)\rangle,
\end{eqnarray}
which is also real. Note the time integral contour of the second term. 

At tree-level, the two-point correlation function, $\langle Q^{I}_{k} Q^{I}_{k} \rangle$, is simply the power spectrum $Q^I_kQ^I_k{}^*$. This  suggests that instead of Wick contracting equation.(\ref{eqn:2pt2}) into Feynman propagators, we should contract them into Wightman functions instead. Let us see how this works by first defining the contraction 
\begin{equation}
\overline{\Q{k}{I}\Q{p}{J}}\equiv \Q{k}{I}\Q{p}{J}-:\Q{k}{I}\Q{p}{J}: ,\label{eqn:wickcontraction}
\end{equation}
where $:\Q{k}{I}\Q{p}{J}:$ is the usual normal ordered product and
\begin{equation}
\int d^3k~\Q{k}{I} = Q^I(\bdv{x},t) = \int d^3 k\,e^{i\bdv{k}\cdot\bdv{x}} \left[a_I(\bdv{k})\U{k}{I}(t)+a_I^{*}(-\bdv{k})\Ud{k}{I}(t)\right] =  \int d^3k\left(\Qp{k}{I}(t)+\Qm{-k}{I}(t)\right). \label{eqn:Qexpansion2}
\end{equation}
The propagator is now \cite{Seery:2008qj}
\begin{equation}
\langle\Q{k}{I}\Q{p}{J}\rangle  = \langle[\Qp{k}{I},\Qm{p}{J}]\rangle  = \U{k}{I}\Ud{p}{J}\delta^I_J\delta^3(\bdv{k}+\bdv{p}) \label{eqn:wightman}.
\end{equation}
With the contraction, equation (\ref{eqn:wickcontraction}), it is straightforward to prove that the correlations of equations (\ref{eqn:2pt2}) and (\ref{eqn:1pt}) are a sum of all possible contractions into both connected and disconnected pieces as per the usual Wick's Theorem in standard quantum field theory. We ignore the disconnected pieces, i.e. the ``vacuum fluctuation'' pieces, in which vertices are connected only to other vertices and no external lines. Since we are computing correlation functions\footnote{Whereas when we compute transition amplitudes, they contribute an overall phase \cite{Peskin:1995ev}.}, the pieces automatically cancel \cite{Weinberg:2005vy}.

As an aside, we note here that in original ``in-in'' formalism of Schwinger-Keldysh (and see also \cite{Jordan:1986ug,Calzetta:1986ey,Collins:2005nu}), the original fields are split into $+$ forward time fields and $-$ backward time fields each with their own generating functional. One can then compute all four possible Green's functions for the fields $(+,+),(-,-),(+,-),(-,+)$, and then all the possible contractions of the correlation will be one of the four above. In other words, the doubling of the fields into $+$ and $-$ sets provides a convenient  book-keeping method for keeping track of  the contours. In our formalism, we have a single contraction equation (\ref{eqn:wickcontraction}), but we pay for this simplicity by with having to explicitly keep track of the contours of the integrals we must perform.

Our strategy is as follows:
\begin{itemize}

\item Expand equation (\ref{eqn:ininformula}) to the desired order keeping careful track of the vacuum prescription contours.

\item{Insert the interaction Hamiltonian $H_{\rm int}$ and expand all fields using equation (\ref{eqn:Qexpansion}).}

\item{Expand using Wick's theorem and the contraction defined in equation (\ref{eqn:wickcontraction}) discarding disconnected diagrams.}

\item{Perform the integral over time(s) leaving only integrals over the internal momenta.}

\item{Regularize the remaining integrals to obtain the final answer.}

\end{itemize}

One should note that these loops contain both UV and IR divergences, unlike the case of \cite{Weinberg:2005vy}. The presence of the IR divergences means that our use of dimensional regularization will yield \emph{incorrect} finite terms. However, we are only interested in the $\log q$ dependence, which is correctly computed by dimensional regularization.  In this paper, we are concerned with large-$N$ effects, but a clear understanding of the mechanism which regulates these logs is clearly of critical importance,   and we plan to pursue this topic in a future publication.\footnote{One can in principle impose a horizon cut-off, as suggested by Lyth in Ref. \cite{Lyth:2007jh} and applied in \cite{Bartolo:2007ti}.  Boyanovsky et al. \cite{Boyanovsky:2004gq, Boyanovsky:2004ph, Boyanovsky:2005sh, Boyanovsky:2005px} have suggested that the IR divergences are regulated by the slow roll limit. However, their approach requires one to analytically continue a combination of the slow roll parameters, which  are physical, and in principle measurable, quantities. }
%-------------------------------------------------------------------------------------------------------------
\subsection{Two-vertex loop}
%-------------------------------------------------------------------------------------------------------------

The two-vertex loop is generated by a three-point interaction term \cite{Seery:2005gb},
\begin{equation}
  H_{\rm int}^{(3)}(t) = \int d^3 x \sum_{I,J}\left[\frac{a^3}{4}\sqrt{2\epsilon_I} Q_I \dot{Q}_J \dot{Q}_J + \frac{a^3}{2} \sqrt{2\epsilon_I}\partial^{-2} \dot{Q}_I \dot{Q}_J \partial^2 Q_J\right],\label{eqn:ijHI}
\end{equation}
with the coupling term, 
\begin{equation}
  \epsilon_I \equiv \frac{1}{2}\frac{\dot{\phi}_I^2}{H^2},
\end{equation}
where here, and in all subsequent calculation in this appendix, we have dropped all factors of $M_p$ to simplify notation. We have also dropped the sums over the $J$ and $I$ since all the fields are identical -- most of the diagrams have identical amplitudes and we will sum them later.

For the two-vertex terms generated by equation (\ref{eqn:ijHI}),  there are two types of vertices,  and we have to compute the contributions from all the possible combinations of two vertices and internal loops (see Fig. (\ref{fig:loopdiagrams}), giving us a total of six diagrams. In the following, we sketch the derivation for the first diagram of Fig. (\ref{fig:loopdiagrams}) with $Q_I\dot{Q}_J\dot{Q}_J$ vertices at both ends -- the other diagrams are computed similarly.
Using the the first term of equation (\ref{eqn:ijHI}) in equation (\ref{eqn:2pt2}), and Wick expanding everything with the contraction of equation (\ref{eqn:wickcontraction}), we get
\begin{eqnarray} \label{eqn:term1int}
&&\int d^3x~e^{i\bdv{q}\cdot(\bdv{x}-\bdv{x}')}\langle{\rm vac, in}| Q^{I}(\bdv{x,\tau})Q^{I}(\bdv{x,\tau}) |{\rm vac, in}\rangle_2 \nonumber \\
&=&-8(2\pi)^9\sum_{J}\Re \left[\int_{-\infty_-}^{\tau}d\tau_2 \int_{-\infty_-}^{\tau_2}d\tau_1 \frac{a^2(\tau_1)\sqrt{2\epsilon_I(\tau_1)}}{4}\frac{a^2(\tau_2)\sqrt{2\epsilon_I(\tau_2)}}{4}\U{q}{I}(\tau_I)\Ud{q}{I}(\tau)\Ud{q}{I}(\tau)\Ud{q}{I}(\tau)\right. \nonumber \\
&&\times\int d^3k \int d^3k'' \dU{k}{J}(\tau_1)\dUd{k}{J}(\tau_2)\dU{k'}{J}(\tau_1)\dUd{k'}{J}(\tau_2)\delta^3(\bdv{q}+\bdv{k}+\bdv{k}') \nonumber \\
&& -\int_{-\infty_+}^{\tau}d\tau_2 \int_{-\infty_-}^{\tau} d\tau_1 \frac{a^2(\tau_1)\sqrt{2\epsilon_I(\tau_1)}}{4}\frac{a^2(\tau_2)\sqrt{2\epsilon_I(\tau_2)}}{4}\U{q}{I}(\tau_1)\Ud{q}{I}(\tau)\Ud{q}{I}(\tau_2)\U{q}{I}(\tau) \nonumber \\
&& \left.\times \int d^3k \int d^3k'' \dU{k}{J}(\tau_1)\dUd{k}{J}(\tau_2)\dUd{k'}{J}(\tau_1)\dUd{k'}{J}(\tau_2)\delta^3(\bdv{q}+\bdv{k}+\bdv{k}') \right],
\end{eqnarray}
where we have dropped all disconnected terms. An overdot in this expression denotes a derivative with respect to conformal time, $\tau$.

We now integrate over the times first, since the interactions of equation (\ref{eqn:ijHI}) satisfy the late time convergence conditions\footnote{See also \cite{vanderMeulen:2007ah}. If we switch the order of integration, we end up swapping an ultraviolet divergence in the integrals over internal momenta for a divergence in $\tau$.}. In near de Sitter space $H\approx \mathrm{const}$, and the mode functions $\U{i}{k}$ can be approximated by
\begin{equation}
\U{k}{I}(\tau) = \sqrt{\frac{H^2}{2(2\pi)^3k^3}}(1+ik\tau)e^{-ik\tau}. \label{eqn:modefunction}
\end{equation}
Although in general the couplings $\epsilon_I(\tau)$ are not constant, since they are changing only slowly we make the further simplifying assumption that they are roughly constant at late times.   Plugging equation (\ref{eqn:modefunction}) into equation (\ref{eqn:term1int}), and then integrating over the times, we obtain
\begin{eqnarray} \label{eqn:term1int2}
&&\int d^3x~e^{i\bdv{q}\cdot(\bdv{x}-\bdv{x}')}\langle {\rm vac, in}| Q^{I}(\bdv{x,\tau})Q^{I}(\bdv{x,\tau}) |{\rm vac, in}\rangle_2 \nonumber \\
&=&\frac{H^4}{16(2\pi)^3}\sum_{J}   \int d^3k~d^3k' \delta^3(\bdv{q}+\bdv{k}+\bdv{k}') ~\left\{\epsilon_I \left[ \frac{5}{4}\frac{kk'}{q^7K}+\frac{3}{4}\frac{kk'}{q^6K^2}++\frac{kk'(K+q)^2}{q^6K^4}\right] \right. \nonumber \\
&+& \left. {\cal{O}}(q\tau) + ~...\right\},
\end{eqnarray}
where we have assumed identical fields hence $\epsilon_I = \epsilon_J$ and used $K\equiv k+k'+q$.

Considering only scales that are larger than the Hubble horizon $q\tau \ll 1$ allows us to also drop the second to last term in equation (\ref{eqn:term1int2}).  The rest of the 2-vertex diagrams can be computed identically, and the final answer we obtain is
\begin{eqnarray} \label{eqn:totalterm2vertex}\nonumber
& & \int d^{3}x\, {\rm e}^{i{\bf q}\cdot({\bf x}-{\bf x'})}\langle {\rm vac, in}|Q^{I}({\bf x},\tau)Q^{I}({\bf x'}, \tau) | {\rm vac, in} \rangle_{2}\\\nonumber   & = & \frac{H^{4}}{2(2\pi)^{3}}N\int d^{3}k\int d^{3}k'\delta({\bf k} + {\bf k'} + {\bf q})\frac{\epsilon_I}{8}\Bigg[\frac{1}{K}\bigg[ \frac{39}{4}\frac{kk'}{ q^{7}}+ \frac{5}{8}\frac{k'}{q^{5}k}- \frac{k'}{q^{4}k^{2}}+ \frac{31}{4}\frac{k'}{q^{3}k^{3}} \bigg] \\\nonumber &&+ \frac{1}{K^{2}}\bigg[\frac{51}{4}\frac{kk'}{q^{6}}+ \frac{31}{2}\frac{ k'}{k^{3}q^{2}} +\frac{5}{8}\frac{k'}{q^{4}k}+ \frac{3}{4}\frac{k'}{q^{3}k^{2}}\bigg] + \frac{1}{K^{3}}\left[ 3\frac{kk'}{q^{5}}+\frac{k'}{q^{2}k^{2}} - \frac{2k'}{qk^{3}} \right] \\ && + \frac{1}{K^{4}}\bigg[\frac{3}{2}\frac{k'}{kq^2} -\frac{1}{2}\frac{kk'}{q^{4}} -2\frac{k'}{k^{2}q}  \bigg] \Bigg]. \end{eqnarray}

Our task now is to compute the physical logs from equation (\ref{eqn:totalterm2vertex}). From simple dimensional analysis, they take the form
\begin{equation}
\int d^3k d^3k'~\delta^3(\bdv{q}+\bdv{k}+\bdv{k}') \frac{k^m k'{}^l}{(q+k+k')^n} \rightarrow q^{3+l+m-n+\delta}F_{(l,m,n)}\ln q + \mathrm{polynomial~divergences}. \label{eqn:dimreg}
\end{equation}
To obtain the coefficient $F_{(l,m,n)}$, we use the identity
\begin{equation}
q\int d^3k d^3k'~\delta^3(\bdv{q}+\bdv{k}+\bdv{k}') f(q,k,k') = 2\pi\int_0^{\infty}k dk \int^{k+q}_{|k-q|}k' dk'~f(q,k,k'). \label{eqn:neatID}
\end{equation}
We can extract the coefficients of physical logs from the above expression using l'H\^{o}pital's rule. We find the limit for both terms going to infinity by differentiating the right hand side of the above identity and equation (\ref{eqn:dimreg})  as many times as needed to tease out the $\log$ divergences, and comparing the result to the same operation applied to the right hand side of equation (\ref{eqn:dimreg}). 

Dropping all other polynomially divergent terms and the $\log$ IR divergences, each term in equation (\ref{eqn:totalterm2vertex}) contribute the following:
\begin{eqnarray} \nonumber
\begin{array}{llllllll}
F_{(1,1,1)} & = \frac{-\pi}{15}, & F_{(1,1,2)} &=\frac{\pi}{3}, & F_{(1,-3,2)} & =\pi, & F_{(1,-3,3)} & = 2\pi,\\
F_{(1,-1,4)}& =0, & F_{(1,-2,2)}&=\pi, & F_{(1,1,4)}&=\frac{\pi}{2}, & F_{(1,1,3)}&=\frac{2\pi}{3},\\
F_{(1,-2,3)}&=0, & F_{(1,-1,1)}&=\frac{-2\pi}{3}, & F_{(1,-2,1)}&=\pi, & F_{(1,-3,1)}&=0,\\F_{(1,-1,2)}&=\pi.
\end{array}
\end{eqnarray}
Given identical fields, such that $\epsilon_I = \epsilon_J$, then putting together everything give us the total contribution from the two-vertex one-loop diagram:
\begin{equation} \label{eqn:term1result}
\int d^3x~e^{i\bdv{q}\cdot(\bdv{x}-\bdv{x}')}\langle {\rm vac, in}| Q^{I}(\bdv{x,\tau})Q^{I}(\bdv{x,\tau}) |{\rm vac, in}\rangle_2 = \frac{H^4 N \epsilon_I}{2(2\pi)^3q^3}\left[\left(\frac{2017}{120}\right)\pi \ln q\right] +...,
\end{equation}
where `$...$' denotes scale-free polynomial divergences.

%-------------------------------------------------------------------------------------------------------------
\subsection{One-vertex loop}
%-------------------------------------------------------------------------------------------------------------

On the other hand, the one-vertex loop  is generated by a four-point interaction term derived in Appendix \ref{App:4ptderivation},
  \begin{eqnarray}
  H_{\rm int}^{(4)}(t) & =&  \int d^3x a^{3}\sum_{I,J}\left[ \frac{1}{4Ha^{2}}\partial_i Q_J\partial_iQ_{J}\partial^{-2}(\partial_j \dot{Q}_I\partial_jQ_I + \dot{Q}_I\partial^2 Q_I)  \right. \nonumber \\
  && +\frac{1}{4H}\dot{Q}_J \dot{Q}_J \partial^{-2}(\partial_i \dot{Q}_I \partial_j Q_I + \dot{Q}_I \partial^2Q_I) \nonumber \\\nonumber
  && + \frac{3}{4H}\partial^{-2}(\partial_j \dot{Q}_J \partial_j Q_{J} + \dot{Q}_J\partial^2 Q_{J})\partial^{-2}(\partial_j \dot{Q}_I \partial_j Q_I + \dot{Q}_I \partial^2 Q_I) \\
  &&\left. +\frac{1}{4}\beta_{2,j}\partial^2 \beta_{2,j}+\dot{Q}_I\beta_{2,i}\partial_i Q_I\right],
  \label{eqn:ijHI2}
  \end{eqnarray}
where,
\begin{eqnarray} \label{eqn:beta2j1}
\frac{1}{2}\beta_{2,j}\simeq \partial^{-4} \left( \partial_{j}\partial_{k}\dot{Q}^{I}\partial_{k}Q_{I} + \partial_{j}\dot{Q}^{I}\partial^{2}Q_{I}- \partial^{2}\dot{Q}^{I}\partial_{j}Q_{I}-\partial_{m}\dot{Q}^{I}\partial_{j}\partial_{m}Q_{I}\right). \end{eqnarray}
The four-point interaction (\ref{eqn:ijHI2}) is explicitly \emph{independent} of the background potential\footnote{The dynamics of the field depends on the potentials and their mode functions, i.e. their green's functions will differ, but the coupling terms have the same structure.} $V$. This means that the one-vertex one-loop correction to a field $I$ from all other fields $J\neq I$ is the same for whether or not $J$ are spectators or participating fields. Each four-point term has the form $Q^I Q^I Q^J Q^J$, meaning that depending on the external lines, we can contract it with either the $J$ or $I$ fields, thus each term will effectively generate two different diagrams. 

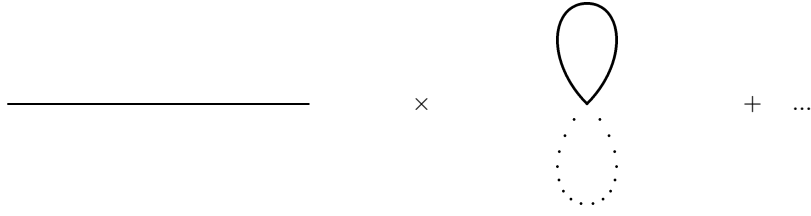
\begin{figure}
\begin{fmffile}{fig3}
\unitlength =1mm
\begin{eqnarray}
&&\langle Q_I(x) Q_I(x')  \partial^{-2}(Q_I(z) Q_I(z)) Q_J(z) Q_J(z)\rangle \propto  \\
{}\nonumber \\
{}\nonumber \\
{}\nonumber \\
&&\parbox{40mm}{
\begin{fmfgraph*}(40,25)
\fmfleft{in}
\fmfright{out}
\fmf{plain}{in,v1,out}
\fmflabel{$I,x$}{in}
\fmflabel{$I,x'$}{out}
\end{fmfgraph*}}
~~~~~~~~~~~~\times
\parbox{40mm}{
\begin{fmfgraph*}(40,25)
\fmfleft{in2}
\fmfright{out2}
\fmf{phantom}{in2,v2,out2}
\fmf{plain,label=$I$,tenstion=1}{v2,v2}
\fmf{dots,label=$J$, left=90,tenstion=1}{v2,v2}
\fmflabel{$z$}{v2}
\end{fmfgraph*}}
+~~~...
\nonumber
\end{eqnarray}
\end{fmffile}

\caption{The non-self-interaction four-point loops factor into a vacuum fluctuation piece times a propagator. The ellipses indicate contractions which lead to polynomial divergences.}
\label{fig:momentumstructure}
\end{figure}

Fortunately,  only the self-interaction term, i.e. $I=J$ contracted with $I$ external lines, contributes a physical log. In the UV, all other terms diverge polynomially and we assume that they can be absorbed by renormalization. Heuristically, this is because the interactions are secretly mediated by gravitons and this, combined with the symmetry of the action, prevents any non-self-interaction loops from contributing, as   described in Section \ref{sect:loop}.  From a diagrammatic perspective,  to yield a log divergence, the final momenta integrand must possess a scale. That is, it must have the form $\sim k^{\alpha}/(\bdv{k}\pm\bdv{q})^{\beta}$ for some $\alpha \in \mathbb{Q}$ and $\beta \in \mathbb{Z}^{+}$, where $\bdv{q}$ and $\bdv{k}$ are the external and internal momenta respectively. However, note that the fields operated by $\partial^{-2}$ is always identically paired i.e. they appear as $\partial^{-2}(Q^{I} Q^{I})$ and never $\partial^{-2}(Q^{I}Q^{J})$. Hence if $I\neq J$, the integrand can only possess momentum factors like $1/({\bf{k}}^2)$ or $1/({\bf{q}}+{\bf{q}}')^2$, which only contribute polynomial divergences. For example, a four-point term with an interaction (dropping time derivatives as they do not affect the final result)  $Q_J \partial^{-2} (Q_I Q_I) Q_J$ for a $\langle Q^I Q^I \rangle$ correlator will yield the diagrams shown in Fig.\ref{fig:momentumstructure} if $I\neq J$, which is simply a vacuum fluctuation diagram multiplied by a propagator. 

We now turn our attention to the self-interaction term, where $I=J$ interaction terms are contracted with $I$ external lines. This calculation is operationally the same as that done by Seery \cite{Seery:2007we} for the single scalar field case using different techniques.  Fourier transforming equation (\ref{eqn:ijHI2}) and switching to conformal time, the interaction Hamiltonian $H_{int}^{(4)}$ becomes
\begin{eqnarray}\nonumber
H_{\rm int}^{(4)} &= & (2\pi)^{3}\int_{-\infty}^{\tau}d\tau'\sum_{I,J}\Bigg[
\frac{a}{4H}\frac{\sigma({\bf k}, {\bf p})}{({\bf k}+{\bf p})^{2}}\dot{Q}_{\bf k}^{I}Q_{\bf p}^{I}\dot{Q}_{\bf a}^{J}\dot{Q}_{\bf b}^{J}\delta({\bf k}+{\bf p}+{\bf a}+{\bf b})
-\frac{a}{4H}{\bf a}\cdot{\bf b}\frac{\sigma({\bf k}, {\bf p})}{({\bf k}+{\bf p})^{2}}\dot{Q}_{\bf k}^{I}Q_{\bf p}^{I}Q_{\bf a}^{J}Q_{\bf b}^{J}\delta({\bf k}+{\bf p}+{\bf a}+{\bf b})
\\&&
+a^{2}\left(\frac{3}{4}\frac{\sigma({\bf k},{\bf p})}{({\bf p} + {\bf k})^{2}}\frac{\sigma({\bf a},{\bf b})}{({\bf a} + {\bf b})^{2}}+\frac{{\bf z}({\bf k}, {\bf p})\cdot{\bf z}({\bf a}, {\bf b})}{({\bf k}+{\bf p})^{4}({\bf a}+{\bf b})^{2}}+2\frac{{\bf z}({\bf k}, \bf{p})\cdot{b}}{({\bf k}+\bf{p})^{4}}\right) \dot{Q}_{\bf k}^{I}Q_{\bf p}^{I}\dot{Q}_{\bf a}^{J}Q_{\bf b}^{J}\delta({\bf k}+{\bf p}+{\bf a}+{\bf b})\Bigg],
\end{eqnarray}
where overdot again denotes derivative with respect to conformal time. We use the notation of \cite{Seery:2007we},
\begin{eqnarray}
{\bf z}({\bf k}, {\bf p})& =&  \sigma({\bf k}, {\bf p})\bf{k} - \sigma({\bf p}, {\bf k}){\bf p}, \\
\sigma({\bf k}, {\bf p}) & = & {\bf k}\cdot{\bf p} + {\bf p}\cdot{\bf p}.
\end{eqnarray}

After some tedious but straightforward calculation reminiscent of the previous section, and considering modes well outside the horizon, $q\tau\rightarrow 0$, we find that the one-vertex self-correction is
\begin{equation} \label{eqn:4ptbigmomint}
\langle Q^{I}(\tau)Q^{I}(\tau) \rangle_{1}= (2\pi)^3 \int d^3\bdv{k}\left[A_s + B_s + C_s \right],
\end{equation}
with the following contributions
\begin{eqnarray}
A_s & = & -\frac{H^{4}}{32(2\pi)^{6}q^{7}} {\bf q}\cdot{\bf k} \Bigg[\left(\frac{6q^2}{k^3}-\frac{5 }{k}\right)\frac{k^{2}+{\bf q}\cdot{\bf k}}{({\bf k}+{\bf q})^{2}}+
\frac{10}{k}\frac{q^{2}+{\bf k}\cdot {\bf q}}{({\bf k}+{\bf q})^{2}}\Bigg],
\end{eqnarray}
\begin{eqnarray}
B_s & = & -\frac{H^{4}}{32(2\pi)^{6}q^{5}}\Bigg[\frac{1}{k}\frac{k^{2}+{\bf q}\cdot {\bf k}}{({\bf k}+{\bf q})^{2}}+\frac{ 5k}{q^2}\frac{q^{2}+{\bf k}\cdot {\bf q}}{({\bf k}+{\bf q})^{2}}\Bigg],
\end{eqnarray}
and
\begin{eqnarray}
C_{s} & = &- \frac{H^{4}}{8(2\pi)^{6}q^{5}}\Bigg[
 \,\frac{5k}{q^{2}}f({\bf k}, {\bf q}, -{\bf k}, -{\bf q})  +\frac{\left(2q^{2}- k^2\right)}{k^{3}}f({\bf q}, {\bf k}, -{\bf q}, -{\bf k})+\frac{6}{k}f({\bf k}, {\bf q}, -{\bf q}, -{\bf k})\Bigg],
\end{eqnarray}
where 
\begin{eqnarray}
f({\bf k},{\bf p}, {\bf a}, {\bf b}) & = & \left(\frac{3}{4}\frac{\sigma({\bf k},{\bf p})}{({\bf p} + {\bf k})^{2}}\frac{\sigma({\bf a},{\bf b})}{({\bf a} + {\bf b})^{2}}+\frac{{\bf z}({\bf k}, {\bf p})\cdot{\bf z}({\bf a}, {\bf b})}{({\bf k}+{\bf p})^{4}({\bf a}+{\bf b})^{2}}+\frac{{\bf z}({\bf k}, \bf{p})\cdot{b}}{({\bf k}+\bf{p})^{4}} +\frac{{\bf z}({\bf a}, \bf{b})\cdot{p}}{({\bf k}+\bf{p})^{4}}\right)\delta({\bf k}+{\bf p}+{\bf a}+{\bf b}).
\end{eqnarray}

The integral in  equation (\ref{eqn:4ptbigmomint}) is divergent and so it needs to be regularized. After regularization we obtain
\begin{equation}
A_s = -\frac{H^4}{2(2\pi)^3q^3}\frac{5\pi}{16} \ln q~,~B_s = -\frac{H^4}{2(2\pi)^3q^3}\frac{\pi}{48} \ln q~,~C_s=-\frac{H^4}{2(2\pi)^3q^3}\frac{\pi}{3} \ln q,
\end{equation}
yielding the following $\log q$ contribution
\begin{equation}
\langle Q^{I}(\tau)Q^{I}(\tau) \rangle = -\frac{H^4}{2(2\pi)^3q^3}\frac{2\pi}{3}\ln q + ... ,\label{eqn:4ptselfcorrection}
\end{equation}
which is our final answer.

Finally,   we sketch the technique we used to extract the $\log q$ terms from equation (\ref{eqn:4ptbigmomint}). The key idea is to write the integrals like
 \begin{equation}
\int d^3 \bdv{k} \frac{\{1,k^i,k^ik^j\}}{k^{2\alpha}({\bf k}\pm{\bf q})^{2\beta}},
\end{equation}
(where $\alpha$ is a half integer and $\beta \in \{1,2,3\}$) as a  sum of terms of the form 
\begin{equation}
f(q^i)\int d^3 \bdv{k} \frac{1}{k^{2\alpha-n}({\bf k}\pm{\bf q})^{2\beta-m}}.\label{eqn:peskinform}
\end{equation}
We can then use standard techniques (e.g. \cite{Peskin:1995ev}) to evaluate the integrals, in combination with the following trick. Define
\begin{eqnarray}
\langle\{1,k^i,k^ik^j\}\rangle_{\alpha, \beta} & \equiv & \int d^d k \frac{\{1,k^i,k^ik^j\}}{(k^2)^{\alpha}({\bf k}\pm{\bf q})^{2\beta}},\\
\langle 1 \rangle_{\alpha, \beta} &= & I_{\alpha, \beta}. \label{eqn:trickint}
\end{eqnarray}
Now since $k^i$ is integrated out, the only vector quantity left is $q^i$ so the following must be true
\begin{equation}
\langle k^i \rangle_{\alpha, \beta} = B_{\alpha, \beta\,}q^i~,~\langle k^i k^j \rangle_{\alpha, \beta} = C_{\alpha, \beta}\,q^i q^j + D_{\alpha, \beta}\, \delta^{ij}q^2, \label{eqn:trickint2}
\end{equation}
where $B_{\alpha, \beta}$, $C_{\alpha, \beta}$ and $D_{\alpha, \beta}$ are coefficients that may contain ultra-violet divergent components.
We can then dot both sides of equation (\ref{eqn:trickint2}) with $q$'s and complete the square to eliminate the numerator $(\qdp)$ terms. For example (for the $B$ term),
\begin{equation}
B_{\alpha, \beta}\,q^2 = \pm\frac{1}{2}\langle (\bdv{q}\pm \bdv{k})^2 \rangle_{\alpha, \beta} \mp \frac{1}{2} \langle k^2 + q^2 \rangle_{\alpha,\beta},
\end{equation}
and the first term cancels one of the powers of $({\bf k}\pm{\bf q})$  while the second term cancels out one power of $k^{2}$. This leaves us with 
\begin{equation}
B_{\alpha,\beta} =\pm \frac{1}{2q^{2}} I_{\alpha, \beta-1} \mp\frac{1}{2}I_{\alpha,\beta} \mp \frac{1}{2q^{2}}I_{\alpha-1,\beta},
\end{equation}
We can iterate this trick until we remove all terms with powers of $({\bf q}\cdot{\bf k})$ in the numerator or cancel all the $({\bf k}\pm{\bf q})$ terms in the denominator, resulting in integrals that are polynomially divergent and hence can be discarded. The remaining integrals are in the form of equation (\ref{eqn:peskinform}) and can be easily regularized. The $C_{\alpha, \beta}$ and $D_{\alpha, \beta}$ terms can be similarly computed by dotting twice with $q$'s.

%-------------------------------------------------------------------------------------------------------------
%-------------------------------------------------------------------------------------------------------------
\section{Multifield 3-point and 4-point Action} \label{App:4ptderivation}
%-------------------------------------------------------------------------------------------------------------
%-------------------------------------------------------------------------------------------------------------

In this paper, we make use of the Arnowitt-Deser-Misner (ADM) formalism \cite{Arnowitt:1962hi} to expand the action equation (\ref{eqn:Nflationaction}) to 4th order in perturbations. The derivation is straightforward if rather tedious (see for example refs. \cite{Seery:2005gb,Seery:2007we} for a detailed application of this formalism), so we simply collect the results.
The background $N$-field action is
\begin{equation}
S = \int d^4 x \sqrt{g}\left[\frac{R}{2} + \sum_I \left( -\frac12 (\partial \phi_I)^2 + V_I(\phi_I)\right)\right],
\end{equation}
the ADM metric is
\begin{equation}
ds^2 = -N^2dt^2 + h_{ij}(dx^i + N^idt)(dx^j + N^j dt),
\end{equation}
and we choose to work in the spatially flat gauge so $h_{ij} = a^2(t)\delta_{ij}$. In other words our metric perturbation has been set to zero by a gauge choice. In addition, we focus on the scalar perturbations, and ignore vector and tensor pieces. This means that the diagrams we computed do not have graviton propagators or loops.
In this gauge the fields have non-zero perturbation
\begin{equation}
\phi_i \rightarrow \phi_I + Q_I.
\end{equation}
Field indices are summed over when contracted, $X^I Y_I = \sum_I X_I Y_I$ and $V = \sum_I V_I(\phi_I)$, with no cross-coupling terms between the fields.

The quadratic action is (exactly) \cite{Seery:2005gb}
\begin{equation}\label{eqn:quadraticaction}
S_2 =\frac{1}{2}\int dt d^3x\, a^3\left[ \dot{Q}_I^2  - a^{-2}(\partial Q_I)^2 - \left(V_{,IJ} - \frac{1}{a^3}\frac{d}{dt}\left(\frac{a^3}{H}\dot{\phi}_I\dot{\phi}_J\right)\right)Q^I Q^J\right].
\end{equation}
The third order action is, to leading order in slow roll \cite{Seery:2005gb},
\begin{eqnarray}\nonumber
S_3 &=& \int dt d^3 x\, a^3 \left[-\frac{1}{4H}\dot{\phi}^JQ_J\dot{Q}^I\dot{Q}_I - \frac{1}{2H}\dot{\phi}^J\partial^{-2}\dot{Q}_J\dot{Q}^I\partial^2Q_I \right. \\
&&+ \left.  \left. \frac{1}{a^3}\frac{\delta L}{\delta Q^J}\right|_1 \left(\frac{\dot{\phi}^J}{4H}\partial^{-2}(Q^I\partial^2 Q_I) - \frac{\dot{\phi}^J}{8H}Q^IQ_I\right)\right],
\end{eqnarray}
where the $\delta L/\delta Q^J $ is the first order equation of motion.

Finally, the fourth order action is, to leading order in slow roll,
\begin{eqnarray}\nonumber S_{4} & = & \int dtd^{3}x\,a^{3}\frac{}{}\left[\frac{3}{4}\left(\partial^{-2}\left( \partial_{j}\dot{Q}^{I}\partial_{j}Q_{I} + \dot{Q}^{I}\partial^{2}Q_{I}\right)\right)^{2} \right. \\  && \left.\frac{}{} -\frac{1}{4}\beta_{2,j}\partial^{2}\beta_{2,j}+\chi_{2}(\partial_{i}\dot{Q}^{I}\partial_{i}Q_{I} +\dot{Q}^{I}\partial^{2}Q_{I}) -\dot{Q}^{I}\beta_{i,2}\partial_{i}Q_{I}\right],  \label{eqn:big4pt}
\end{eqnarray} 
where the auxiliary fields $\chi_2$ and $\beta_{2,j}$ are, to leading order,
\begin{eqnarray}\label{eqn:beta2j}
\frac{1}{2}\beta_{2,j}\simeq \partial^{-4} \left( \partial_{j}\partial_{k}\dot{Q}^{I}\partial_{k}Q_{I} + \partial_{j}\dot{Q}^{I}\partial^{2}Q_{I}- \partial^{2}\dot{Q}^{I}\partial_{j}Q_{I}-\partial_{m}\dot{Q}^{I}\partial_{j}\partial_{m}Q_{I}\right), \end{eqnarray}
\begin{equation}
\partial^{2}\chi_{2} = -\frac{1}{4a^{2}H}\partial_{i}Q^{I}\partial_{i}Q_{I } - \frac{3}{2}\partial^{-2}\left( \partial_{j}\dot{Q}^{I}\partial_{j}Q_{I} + \dot{Q}^{I}\partial^{2}Q_{I}\right)-\frac{1}{4H}\dot{Q}^{I}\dot{Q}_{I}.
\end{equation}
Note that the 4-point action equation (\ref{eqn:big4pt}) will generate $N^2$ non-trivial diagrams, since there are two sums over the field indices. Note also here that this expression reduces to those of \cite{Seery:2007we} in the single field limit.

%-------------------------------------------------------------------------------------------------------------
%---------------------------------------------------------------------------------------------------
 \section{Quantization of Theories with Derivative Interactions} \label{app:canquant}
 %---------------------------------------------------------------------------------------------------
%-------------------------------------------------------------------------------------------------------------

 In this appendix, we describe the procedure we use to canonically quantize the classical theory, as we encounter Lagrangians with interactions containing time-derivatives of the fields. A path integral formalism is given in \cite{Weinberg:1995mt}. For theories with up to 2nd order in time derivatives, a treatment is given in \cite{Gerstein:1971fm}.  In the problem we are considering, we encounter interactions up to 3rd order in time derivatives, so we extend \cite{Gerstein:1971fm} at least to the next to leading order in slow-roll. Extension to all orders is straightforward which we will leave for future work. A path-integral approach is pursued by Seery \cite{Seery:2007we}.

 We follow the usual procedure in canonically quantizing a classical theory specified by a Lagrangian density,  $\mathcal{L}(Q, \dot{Q})$. That is, we define the momenta conjugate to the field $Q$ by;
 \begin{eqnarray}
 \pi = \frac{\partial \mathcal{L}}{\partial \dot{Q}},
 \end{eqnarray}
 and construct the Hamiltonian density, $\mathcal{H}$, as the Legendre transform of the Lagrangian density;
 \begin{eqnarray}
 \mathcal{H} & = & \pi\dot{Q} - \mathcal{L},
 \end{eqnarray}
 where $\dot{Q}$ is expressed in terms of $\pi$. We then move to an interaction picture by separating the Hamiltonian into its quadratic part $\mathcal{H}_{0}$ and higher order part $\mathcal{H}_{\rm int}$ and replace $\pi$ in $\mathcal{H}_{\rm int}$ with the interaction picture $\pi_{I}$ given by
 \begin{eqnarray}
 \dot{Q} = \pi_{I} & = & \left.\frac{\partial \mathcal{H}_{0}}{\partial \pi}\right|_{\pi = \pi_{I}}.
 \end{eqnarray}

 The question we want to address here is, what is $\mathcal{H}_{\rm int}$? Naively, one might guess that $\mathcal{H}_{\rm int} = -\mathcal{L}_{\rm int}$, where $\mathcal{L}_{\rm int} = \mathcal{L}-\mathcal{L}_{0}$ and $\mathcal{L}_{0}$ is the quadratic part of $\mathcal{L}$. If the only time derivatives of the field are in a canonical kinetic term, this is certainly the case. However, when time derivatives are present in the interaction terms, these can modify the relation between $\pi$ and $\dot{Q}$, and the construction of the Hamiltonian then generates extra interactions. Fortunately, at the order we are working the additional terms generated are either subleading in slow roll, or higher order in the fluctuations. To see this, note that the Lagrangians we consider have the schematic form
 \begin{eqnarray}
 \mathcal{L} & = &  \frac{1}{2}\dot{Q}^{2}-V(Q) + \left(\sqrt{\epsilon} f_{2}+f_{3}\right) \dot{Q} + \frac{1}{2}\left(\sqrt{\epsilon} g_{1}+g_{2}\right)\dot{Q}^{2}+\frac{1}{3}h_{1}\dot{Q}^{3} + \mathcal{O}(\epsilon Q^{3})+ \mathcal{O}(\epsilon Q^{4}) + \mathcal{O}(Q^{5})
 \end{eqnarray}
 where $\epsilon$ is the usual slow roll epsilon and the subscripts of $f_{m}$ and $g_m$ denote a term of order $m$ in fluctuations, $Q$. The terms containing no time derivatives are gathered into $V(Q)$. To proceed, we assume that $|Q|\sim|\dot{Q}|\sim|\pi|$. A straightforward calculation then shows that
 \begin{eqnarray}
 \mathcal{H}   & = & \frac{\dot{Q}^{2}}{2}+V(Q) - \left(\sqrt{\epsilon} f_{2} + f_{3}\right)\dot{Q} - \frac{1}{2}\left(\sqrt{\epsilon} g_{1}+g_{2}\right)\dot{Q}^{2} - \frac{1}{3}h_{1}\dot{Q}^{3}+\epsilon\left(f_{2}g_{1}\dot{Q} + \frac{1}{2} g_{1}^{2}\dot{Q}^{2}\right)\\\nonumber
 && +\mathcal{O}(\epsilon\, Q^{3})+ \mathcal{O}(\epsilon\, Q^{4}) + \mathcal{O}(Q^{5}).\\
 & = & \mathcal{H}_{0}-\mathcal{L}_{I}+\mathcal{O}(\epsilon\, Q^{3})+ \mathcal{O}(\epsilon\, Q^{4}) + \mathcal{O}(Q^{5}).
 \end{eqnarray}
 So, to leading order in slow roll and to fourth order in fluctuations, it is safe to take $\mathcal{H}_{\rm int} = -\mathcal{L}_{\rm int}$, the correction being at the most of ${\cal O}(\epsilon)$ and thus subleading during inflation.

\bibliography{nfield}

\end{document}